\documentclass{article}
\usepackage{graphicx}
\usepackage{booktabs}
\usepackage{amsmath}
\usepackage{hyperref}
\usepackage{geometry}
\usepackage{amsmath}
\usepackage{enumitem}
\usepackage{array}
\usepackage{booktabs} 
\usepackage{float}
\title{Economic Causal Inference Based on DML Framework: Python Implementation of Binary and Continuous Treatment Variables}
\author{Shunxin Yao \\QUT\\ n12012416@qut.edu.au}
\date{}

\begin{document}

\maketitle

\begin{abstract}
This study utilizes a simulated dataset to establish Python code for Double Machine Learning (DML) using Anaconda's Jupyter Notebook and the DML software package from GitHub. The research focuses on causal inference experiments for both binary and continuous treatment variables. The findings reveal that the DML model demonstrates relatively stable performance in calculating the Average Treatment Effect (ATE) and its robustness metrics. However, the study also highlights that the computation of Conditional Average Treatment Effect (CATE) remains a significant challenge for future DML modeling, particularly in the context of continuous treatment variables. This underscores the need for further research and development in this area to enhance the model's applicability and accuracy.

\end{abstract}

\section{Introduction}
The background of causal inference in economics can be traced back to the need for a deep understanding of economic phenomena and their underlying mechanisms, aiming to reveal the impact of one variable (the treatment variable) on another variable (the outcome variable). This is particularly crucial in policy evaluation, experimental design, and economic model construction. With advancements in econometrics and statistics, economists have recognized that relying solely on correlation cannot reveal causality, making causal inference methods an essential part of economic research. In the framework of causal inference, the nature of the treatment variable influences the analysis results; discrete treatment variables typically have finite values, such as policy acceptance and program participation. The main challenge here is how to control for potential confounding factors and avoid selection bias affecting causal effect estimates. Analysis often relies on methods like instrumental variables and propensity score matching. In contrast, continuous treatment variables can take any value, such as policy intensity and tax rates, which may lead to nonlinear relationships. Therefore, complex models with more interaction terms and nonlinear components are required in causal inference to capture the heterogeneity of intervention effects (Pearl, Glymour, \& Jewell, 2016).

These challenges highlight the complexity of economic research, and researchers must choose appropriate methods based on specific issues and data characteristics to ensure the effectiveness of causal inference. Double Machine Learning (DML) framework can handle the problem. By integrating traditional economic models with modern machine learning techniques, the bias issues in causal effect estimation can be effectively addressed, and stepwise estimation is used to improve accuracy. DML consists of two main steps: first, using machine learning algorithms to model confounding factors and eliminate their influence; then, after controlling for these confounding factors, using machine learning tools to estimate the direct impact of treatment variables on the dependent variable (Chernozhukov et al., 2018). The advantages of DML lie in enhancing estimation accuracy, handling high-dimensional data, and strengthening the robustness of causal inference, making it particularly suitable for estimating causal effects in complex data environments. This paper aims to demonstrate how to apply the DML framework to handle causal inference with discrete and continuous treatment variables using Python code.

\section{Literature Review}
In economics, causal inference is a crucial tool for understanding relationships between variables. Common methods of causal inference include regression discontinuity design (RDD), instrumental variable (IV) methods, and propensity score matching (PSM). RDD divides the sample into treatment and control groups using a "threshold," allowing for an estimation of causal effects by comparing outcomes near the threshold. This method is suitable for evaluating policy or intervention effects. The instrumental variable approach addresses endogeneity issues by introducing a tool variable that correlates with the independent variable but not with the error term of the dependent variable, thus providing more reliable estimates of causal effects.

Propensity score matching calculates the probability of everyone receiving the treatment to match the treatment and control groups, aiming to control for observed confounding variables to reduce selection bias and estimate treatment effects (Pearl, Glymour, \& Jewell, 2016). Among these traditional causal inference methods, the double machine learning (DML) framework, as an emerging tool, combines the advantages of new machine learning techniques and traditional statistical methods, offering a more flexible, effective, and robust approach to causal inference. DML uses machine learning algorithms to model potential confounding factors flexibly, avoiding the need for strict assumptions about relationships between variables, effectively handling high-dimensional data and nonlinear relationships. In this way, DML can address complex data structures and when dealing with potential confounding factors, it demonstrates its unique advantages, not only improving the efficiency of causal effect estimation but also providing consistent estimation results even when model specifications are incomplete, or selection bias exists. Moreover, the DML framework can be combined with other traditional causal inference methods such as IV and RDD, thereby enhancing the depth and breadth of causal inference (Chernozhukov et al., 2018).

Propensity Score Matching (PSM) and Instrumental Variable (IV) were two commonly used methods in causal inference. PSM matches treatment and control groups by estimating the probability of individuals being treated, reducing selection bias. However, it relies on the accuracy of the model, cannot control unobservable confounding factors, and may degrade matching quality in high-dimensional data. The IV method addresses endogeneity by introducing instrumental variables that are correlated with the treatment variable but only with the outcome variable through the treatment variable. However, selecting instrumental variables and meeting exogeneity assumptions can be challenging, and weak instrument problems may affect the validity of estimates. Therefore, despite their significant applications in controlling confounding and endogeneity, researchers must carefully consider the limitations of PSM and IV. To address these shortcomings of traditional methods, DML models have emerged (Fuhr, Berens, \& Papies, 2024).

Chernozhukov et al. (2018) proposed the Double Machine Learning (DML) method, which is used to estimate treatment effects and structural parameters in causal inference for high-dimensional data. By combining machine learning with semi-parametric methods, DML can eliminate model bias, providing more accurate and unbiased estimates. They also demonstrated the application of this method in econometrics, particularly in addressing parameter estimation issues in policy evaluation and economic models. Building on this foundation, Chernozhukov et al. (2024) further advanced the Double Machine Learning (DML) method, which combines the orthogonality of Neyman and cross-fitting to reduce the impact of high-dimensional interference parameter estimation errors on causal inference. This approach achieves \(\sqrt{\text{N}}\) convergence in target parameter estimation while maintaining statistical inference validity. DML is applicable to partial linear regression, instrumental variable models, and treatment effect estimation, providing a robust methodological foundation for high-dimensional causal inference.

In practice, Flores and Chernozhukov (2018) established an open-source code library for the GitHub project, demonstrating how to implement DML in Python for causal inference and effect estimation. The library provides detailed documentation and examples, integrating machine learning methods such as random forests and Lasso regression, and supports model evaluation and result visualization, making it suitable for causal inference analysis of high-dimensional data. Additionally, Bach et al. (2022) introduced a Python open-source library for double machine learning (DML), designed to achieve robust causal inference through Neyman orthogonality, machine learning methods, and sample partitioning. It adopts object-oriented programming (OOP), supports various causal models (such as partially linear regression and instrumental variable regression), and provides a flexible API for extensible functionality. Compared to EconML and CausalML, DML places greater emphasis on the validity of orthogonality conditions and statistical inference.

\section{Data}
\subsection{Source of Data}
The data is generated through simulation. In the binary treatment setting, the independent variables are generated from a standard normal distribution, forming a 1000 by 3 matrix representing covariates. The treatment variable is created by combining two of the independent variables with added noise, then converting it into a binary variable (0 or 1) to indicate whether treatment is applied. The outcome variable is generated using a linear combination of the independent variables, the binary treatment variable, and random noise, capturing both the treatment effect and variability. In the continuous treatment setting, the independent variables are similarly generated from a standard normal distribution, while the treatment variable is directly sampled from a standard normal distribution to represent treatment intensity or dosage. The outcome variable is also generated using a linear combination of the independent variables, the continuous treatment variable, and random noise. Both settings simulate realistic data structures for causal inference, with the key difference being the binary versus continuous nature of the treatment variable.

\subsection{Feature of Data}
\subsubsection{Binary Treatment of Variables}
\begin{table}[H]
\centering
\caption{Statistics of the Binary Treatment Dataset}
\begin{tabular}{lccccc}
\toprule
 & count & mean & std & min & max \\
\midrule
X1 & 1000 & 0.06 & 0.97 & -2.94 & 3.93 \\
X2 & 1000 & -0.01 & 1.01 & -3.24 & 3.19 \\
X3 & 1000 & 0.05 & 0.98 & -3.02 & 3.85 \\
Y & 1000 & 0.87 & 4.08 & -11.27 & 14.74 \\
D & 1000 & 0.51 & 0.50 & 0.00 & 1.00 \\
\bottomrule
\end{tabular}
\end{table}

\begin{figure}[H]
\centering
\includegraphics[width=0.7\textwidth]{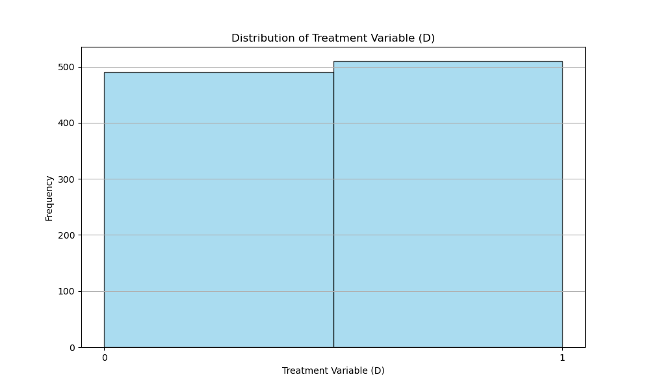}
\caption{Distribution of the Binary Variable}
\end{figure}

In the data of binary treatment variables, the treatment variable D is a binary variable (0 or 1), with a mean of 0.51, indicating a nearly uniform distribution, suggesting that the sample sizes of the treatment group and control group are roughly balanced. The independent variables are $X_1$, $X_2$, and $X_3$. $X_1$, $X_2$, and $X_3$ all conform to a standard normal distribution, with means close to 0 and standard deviations close to 1, indicating reasonable data generation. The mean of the dependent variable Y is 0.869, but the standard deviation is relatively large (4.084), suggesting significant data volatility and potential strong treatment effects or noise. This data is suitable for estimating the average treatment effect (ATE) of discrete treatments.

\subsubsection{Continuous Treatment Variables}
\begin{table}[H]
\centering
\caption{Statistics of the Continuous Treatment Dataset}
\begin{tabular}{lccccc}
\toprule
 & count & mean & std & min & max \\
\midrule
X1 & 1000 & 0.06 & 0.97 & -2.94 & 3.93 \\
X2 & 1000 & -0.01 & 1.01 & -3.24 & 3.19 \\
X3 & 1000 & 0.05 & 0.98 & -3.02 & 3.85 \\
Y & 1000 & 0.08 & 3.92 & -12.37 & 14.23 \\
D & 1000 & -0.02 & 1.03 & -2.93 & 3.24 \\
\bottomrule
\end{tabular}
\end{table}

\begin{figure}[H]
\centering
\includegraphics[width=0.7\textwidth]{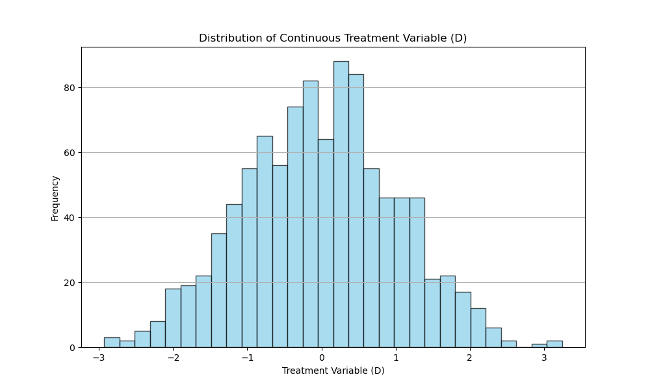}
\caption{Distribution of the Continuous Variable}
\end{figure}

In the data with continuous treatment variables, variable D is a continuous variable, with a mean of -0.019 and a standard deviation of 1.027, conforming to the characteristics of a standard normal distribution. Independent variables are $X_1$, $X_2$, and $X_3$. $X_1$, $X_2$, and $X_3$ also follow a standard normal distribution, with means close to 0 and standard deviations close to 1, indicating reasonable data generation. The dependent variable Y has a mean of 0.076 and a standard deviation of 3.921, showing significant data fluctuation, which may indicate strong treatment effects or noise. This data is suitable for estimating the marginal causal effect of continuous treatment.

\subsubsection{Comparison of Binary and Continuous Treatment Variables}
Binary treatment variables and continuous treatment variables have highly consistent distributions on the independent variable X, both conforming to the standard normal distribution. The main difference lies in the treatment variables D: binary treatment variables are binary, suitable for estimating the average treatment effect (ATE) of discrete treatments; while continuous treatment variables are continuous, suitable for estimating the marginal causal effect of continuous treatments. The dependent variable Y exhibits significant volatility in both cases, indicating that there may be strong treatment effects or noise in the data. Both types of data are suitable for causal inference analysis, but the appropriate models and methods should be chosen based on the type of treatment variable.

\section{Methodology}
\subsection{Theoretical Level}
\subsubsection{The Core Idea of Double Machine Learning (DoubleML)}
The goal of DML is to estimate the causal effect of the treatment variable D on the outcome variable Y while controlling for the influence of covariates X. Its advantage lies in its applicability to high-dimensional data and nonlinear relationships. Additionally, it can eliminate the confounding effects of covariates X through residualization, yielding unbiased estimates of causal effects.

\textbf{Step 1.} Machine learning models (such as RandomForestRegressor) are used to fit separately:
The prediction model of Y for X: \( \hat{Y}(X) \)
The prediction model of D for X: \( \hat{D}(X) \)

\textbf{Step 2.} Calculate the residual:
The residual of Y:
\[ \hat{Y} = Y - \hat{Y}(X) \]
The residual of D:
\[ \hat{D} = D - \hat{D}(X) \]

\textbf{Step 3.} The causal effect \(\theta\) is estimated by regressing \(Y\) on \(\hat{D}\).

\subsubsection{Discrete Treatment Variables (e.g., Binary Treatment)}
The treatment variable \(D_{\text{dummy}}\) is binary (0 or 1) indicating whether the treatment is accepted. The causal effect is estimated by Partial Linear Regression (PLR) model.

\begin{equation}
Y = \theta D_{\text{dummy}} + g(X) + \epsilon
\end{equation}

where:
\(Y\) is the dependent variable.
\(D_{\text{dummy}}\) is the binary treatment variable.
\(X\) is the independent variable.
\(g(X)\) is a nonlinear function of the independent variable.
\(\epsilon\) is the error term.
\(\theta\) is the causal effect (ATE).

\subsubsection{Continuous Treatment Variables (Continuous Treatment)}
Continuous treatment variables are denoted as \(D_{\text{continuous}}\), which indicates the intensity of treatment. The causal effect is estimated by the Partial Linear Regression (PLR) model:

\begin{equation}
Y = \theta D_{\text{continuous}} + g(X) + \epsilon
\end{equation}

where:
\(Y\) is the dependent variable.
\(D_{\text{continuous}}\) is a continuous treatment variable.
\(X\) is the independent variable.
\(g(X)\) is a nonlinear function of the independent variable.
\(\epsilon\) is the error term.
\(\theta\) is the causal effect (ATE).

\subsubsection{Summary from both views}
\begin{table}[H]
\centering
\caption{Holistic Steps}
\begin{tabular}{@{}>{\bfseries}m{3cm} m{10cm}@{}}
\toprule
\textbf{Step} & \textbf{Description} \\
\midrule
Step 1 & Generate simulated data: Create the independent variable \(X\), the binary or continuous treatment variable \(D\), and the dependent variable \(Y\). \\
Step 2 & Use RandomForestRegressor as a machine learning model: Estimate \(g(X)\) using the RandomForestRegressor model. \\
Step 3 & Create a DoubleMLData object: Pass in the data \(X\), \(D\), and \(Y\) into the DoubleMLData object. \\
Step 4 & Estimate causal effects: Estimate the causal effects \(\theta\) using the DoubleMLPLR model. \\
Step 5 & Fit the model and output the estimated causal effect: Estimate the Average Treatment Effect (ATE) and display the results. \\
Step 6 & Obtain standard error and confidence interval: Use the \texttt{summary} method to obtain the standard error and confidence interval of the estimated causal effect. \\
\bottomrule
\end{tabular}
\end{table}

\subsection{Practical Level}
\subsubsection{Libraries and Tools Used}
\begin{table}[H]
\centering
\caption{Libraries and Tools Used}
\begin{tabular}{p{4cm}p{8cm}}
\toprule
Library/tool & Brief introduction \\
\midrule
numpy & Used for efficient array and matrix operations, generating simulated data (such as independent variable X, treatment variable D and dependent variable Y). \\
pandas & Used for data processing and operation, the generated data is organized into DataFrame format for subsequent analysis. \\
sklearn & Machine learning models are provided, and RandomForestRegressor is used as the base learner in the code to fit nonlinear relationships. \\
doubleml & The Double Machine Learning (DML) framework is used to estimate causal effects and supports discrete and continuous treatment of variables. \\
\bottomrule
\end{tabular}
\end{table}

\subsubsection{Binary Data DML Code Steps}
In the code for binary data double machine learning (DML), first, independent variables X are generated from a standard normal distribution, and treatment variable D is processed through linear combinations of \(X[:,0] + X[:,1]\), and \(X[:,0]\), added noise and then converted into binary variables (0 or 1). The dependent variable Y is generated using a linear model: \( Y = 3X_1 + 2X_2 + 1.5D + \epsilon \). Next, RandomForestRegressor is used as the base learner to construct a DoubleMLPLR model, fit the data, and estimate the average treatment effect (ATE). Finally, the causal effect estimates, and their statistical significance are output.

\subsubsection{Continuous Data DML Code Steps}
In the code for double machine learning (DML) with continuous data, first generate independent variables X from a standard normal distribution. The treatment variable D is directly generated as a continuous value from a standard normal distribution. The dependent variable Y is generated through the linear model \( Y = 3X_1 + 2X_2 + 1.5D + \epsilon \). Next, use RandomForestRegressor as the base learner to construct a DoubleMLPLR model, fit the data, and estimate the average treatment effect (ATE). Finally, output the causal effect estimates along with their statistical significance.

\subsubsection{Comparison Summary}
The DML code steps for binary and continuous data are similar, with the main difference lying in how variable D is generated: binary data is transformed into binary variables through logical conditions, while continuous data is directly generated from a normal distribution. Both methods estimate causal effects using a double machine learning framework and output treatment effect's point estimates, standard errors and confidence intervals.

\section{Code Implementation}
\subsection{Binary Treatment}
\begin{verbatim}
import numpy as np
import pandas as pd
from sklearn.ensemble import RandomForestRegressor
import doubleml as dml

# Set random seed for reproducibility
np.random.seed(42)

# Generate simulated data
n_samples = 1000
n_features = 3

# Independent variables X (n_samples x n_features)
X = np.random.randn(n_samples, n_features)

# Treatment variable D (n_samples x 1), binary treatment
D = (X[:, 0] + X[:, 1] + np.random.randn(n_samples)) > 0
D = D.astype(int) # Convert to binary (0 or 1)

# Outcome variable Y (n_samples x 1)
Y = 3 * X[:, 0] + 2 * X[:, 1] + 1.5 * D + np.random.randn(n_samples)

# Create DataFrame
data = pd.DataFrame(X, columns=[fX{i+1}' for i in range(n_features)])
data["Y"] = Y
data["D"] = D

# Set independent and dependent variables
X = data[['X1', 'X2', 'X3']] # Independent variables
y = data["Y"] # Outcome variable
d = data["D"] # Treatment variable

# Define machine learning model
ml_model = RandomForestRegressor(n_estimators=100)

# Create DoubleML dataset
dml_data = dml.DoubleMLData.from_arrays(X.values, d.values, y.values)

# Set up DoubleML model
dml_model = dml.DoubleMLPLR(dml_data, ml_model, ml_model)

# Fit the model
dml_model.fit()

# Output the estimated treatment effect (ATE)
print("Estimated treatment effect (ATE):", dml_model.coef)

# Use summary to get standard errors and confidence intervals
summary = dml_model.summary
print("Model Summary:\n", summary)
\end{verbatim}

\subsection{Continuous Treatment}
\begin{verbatim}
import numpy as np
import pandas as pd
from sklearn.ensemble import RandomForestRegressor
import doubleml as dml

# Set random seed for reproducibility
np.random.seed(42)

# Generate simulated data
n_samples = 1000
n_features = 3

# Independent variables X (n_samples x n_features)
X = np.random.randn(n_samples, n_features)

# Treatment variable D (n_samples x 1), continuous treatment
D = np.random.randn(n_samples) # Continuous treatment variable

# Outcome variable Y (n_samples x 1), including treatment effect
Y = 3 * X[:, 0] + 2 * X[:, 1] + 1.5 * D + np.random.randn(n_samples)

# Create DataFrame
data = pd.DataFrame(X, columns=[fX{i+1}' for i in range(n_features)])
data["Y"] = Y
data["D"] = D

# Set independent and dependent variables
X = data[['X1', 'X2', 'X3']] # Independent variables
y = data["Y"] # Outcome variable
d = data["D"] # Treatment variable

# Define machine learning model
ml_model = RandomForestRegressor(n_estimators=100)

# Create DoubleML dataset
dml_data = dml.DoubleMLData.from_arrays(X.values, d.values, y.values)

# Set up DoubleML model
dml_model = dml.DoubleMLPLR(dml_data, ml_model, ml_model)

# Fit the model
dml_model.fit()

# Output the estimated treatment effect (ATE)
print("Estimated treatment effect (ATE):", dml_model.coef)

# Use summary to get standard errors and confidence intervals
summary = dml_model.summary
print("Model Summary:\n", summary)
\end{verbatim}

\section{Results}
\begin{table}[h!]
\centering
\caption{Results of Models}
\begin{tabular}{lccccc}
\toprule
Model & ATE & std & t-value & p-value & 95\% CI \\
\midrule
Binary Treatment & 0.148218 & 0.008722 & 16.993901 & \(9.11\times 10^{-65}\) & [0.131124, 0.165313] \\
Continuous Treatment & 0.414707 & 0.010496 & 39.50954 & 0.000 & [0.394134, 0.435279] \\
\bottomrule
\end{tabular}
\end{table}

\subsection{Model Results for Binary Variables}
In the model with binary treatment variables, the estimated average treatment effect (ATE) is 0.148, with a standard error of 0.009, a t-value of 16.994 and a p value nearly to 0, indicating that the treatment effect is significant. The 95

\subsection{Model Results for Continuous Treatment Variables}
In the model with continuous treatment variables, the estimated average treatment effect (ATE) is 0.415, with a standard error of 0.010, a t-value of 39.510 and a p-value close to 0, suggesting a significant treatment effect. The 95

\subsection{Comparison of Results for Binary and Continuous Treatment Variables}
The effect size of continuous treatment variables (0.415) is significantly larger than that of binary treatment variables (0.148), indicating a stronger impact of continuous treatment on the dependent variable. The p-values for both models are close to 0, suggesting highly significant statistical effects. The confidence interval for continuous treatment variables is narrower ([0.394,0.435]), indicating higher estimation accuracy, whereas the confidence interval for binary treatment variables is wider ([0.131,0.165]), indicating relatively lower estimation accuracy. Whether discrete or continuous treatment variables, treatment variable D has a significant positive impact on dependent variable Y, but the treatment effect of continuous treatment is larger and more precisely estimated.

\section{Conclusions}
The Double machine learning (DML) framework has demonstrated significant advantages in causal inference, particularly when dealing with discrete and continuous treatment variables. First, DML effectively controls potential confounding factors by combining machine learning methods with traditional statistical inference. This is especially true for discrete treatment variables, where DML can reduce the bias between dependent and independent variables through efficient predictive models, thereby improving estimation accuracy. This is more challenging in regression discontinuity designs (RDD) and instrumental variable (IV) methods, as these typically rely on strict model assumptions that may lead to bias. Second, the flexibility of the DML framework is particularly prominent for continuous treatment variables. Traditional methods often require linear or specific forms of assumptions about treatment effects, whereas DML, using non-parametric or semi-parametric machine learning models, can capture more complex nonlinear relationships, enhancing the effectiveness of estimating heterogeneous treatment effects. Additionally, DML exhibits excellent out-of-sample predictive capabilities, avoiding overfitting through techniques such as cross-validation, which improves the robustness of estimates, especially in policy evaluation and practical applications where the generalization effect is crucial. The scalability and adaptability of DML enable it to integrate with various machine learning algorithms, making it suitable for different types of data and research. To solve this problem, DML has become an important tool for causal inference in the era of big data (Chernozhukov et al., 2018).

However, despite the significant advantages of the DML framework in causal inference, there is still considerable room for future research. First, for conditional average treatment effects (CATE) estimation with continuous treatment variables, research can focus on developing new machine learning algorithms that integrate deep learning techniques to more accurately capture the heterogeneity of treatment effects. Second, model selection and optimization are crucial in the DML framework. Future studies should explore how to select the most suitable machine learning models under different data structures and types of treatment variables and develop automated model selection methods to improve the efficiency and accuracy of causal inference. As data dimensions increase, conducting causal inference in high-dimensional settings becomes a challenge. Future research could concentrate on the application of high-dimensional feature selection and dimensionality reduction techniques in the DML framework to enhance the stability and accuracy of estimates. Additionally, future research should focus on the robustness of the DML framework, developing new robustness testing methods and sensitivity analyses to ensure the reliability of causal inference results. The cross-domain application of the DML framework also holds significant potential. Future research should explore its applicability in fields such as healthcare, economics, and social sciences, developing specific DML methods tailored to the needs of each domain. Finally, in policy evaluation, DML can further integrate randomized controlled trials (RCT), improve the accuracy of external validity tests, and provide a more comprehensive causal effect assessment for policy interventions. Through these improvements, the DML framework will provide stronger support for both the theory and practice of causal inference.

\section*{References}
\begin{itemize}
\item Pearl, J., Glymour, M., \& Jewell, N. P. (2016). Causal inference in statistics: A primer. Wiley.
\item Chernozhukov, V., Chetverikov, D., Demirer, M., Duflo, E., Hansen, C., Newey, W., \& Robins, J. (2018). Double/debiased machine learning for treatment and structural parameters. \textit{The Econometrics Journal}, 21(1), C1--C68.
\item Chernozhukov, V., Chetverikov, D., Demirer, M., Duflo, E., Hansen, C., Newey, W., \& Robins, J. (2024). Double/Debiased Machine Learning for Treatment and Structural Parameters. \textit{arXiv preprint arXiv:1608.00060v7}.
\item Flores, A. B., \& Chernozhukov, V. (2018). DoubleML for Python (Version 0.0.1) [Computer software]. GitHub.
\item Bach, P., Chernozhukov, V., Kurz, M. S., \& Spindler, M. (2022). DoubleML - An object-oriented implementation of double machine learning in Python. \textit{Journal of Machine Learning Research}, 23(2022), 1--6.
\item Fuhr, J., Berens, P., \& Papies, D. (2024). Estimating causal effects with double machine learning - A method evaluation. \textit{School of Business and Economics, University of Tubingen}.
\end{itemize}

\section*{Acknowledgment}
I acknowledge the use of AI tools in the preparation of this research. ChatGPT and DeepSeek were consulted for general suggestions and insights, which were carefully reviewed and refined by the author. Additionally, WPS was used for language assistance to enhance the clarity and readability of the manuscript. All final decisions regarding content, analysis, and conclusions were made by the author.

\end{document}